\documentclass[epj]{svjour}
\usepackage{graphics}
\usepackage{amsmath}
\usepackage{amssymb}
\usepackage{graphicx}
\usepackage{dcolumn}
\usepackage{float}
\usepackage{epsfig,color}
\usepackage{bm}
\usepackage{longtable}

\newcommand{\ba}{\begin{eqnarray}}
\newcommand{\ea}{\end{eqnarray}}
\begin{document}
\title{Self-energies of ground-state octet and decuplet baryon states due to the coupling to the baryon-meson continuum}
\author{H. Garc\'ia-Tecocoatzi\inst{1,2}  \and R. Bijker\inst{2} \and J. Ferretti \inst{3,4}\and 
E. Santopinto\inst{1,}
\thanks{\emph{Corresponding author} santopinto@ge.infn.it}%
}                     
\offprints{}          
\institute{INFN, Sezione di Genova, via Dodecaneso 33, 16146 Genova (Italy). \and Instituto de Ciencias Nucleares, Universidad Nacional Aut\'onoma de M\'exico, 04510 M\'exico DF, M\'exico. \and Institute of Theoretical Physics, Chinese Academy of Sciences. Beijing 100190, China.  \and Dipartimento di Fisica and INFN, Universit\`a di Roma  Sapienza, Piazzale A. Moro 5, 00185, Roma, Italy.}
\date{Received: date / Revised version: date}
%
\abstract{
We present an unquenched quark model calculation of the mass shifts of ground-state octet and decuplet baryons due to the coupling to the meson-baryon continuum. 
The  $q\bar{q}$ pair-creation effects are taken explicitly into account through a microscopic, QCD-inspired, quark-antiquark pair-creation mechanism.
} 
%
\maketitle
\section{Introduction}
Many studies investigate hadron properties within the quark model (QM). The QM \cite{Eichten:1974af,Isgur:1979be,Godfrey:1985xj,Capstick:1986bm,Iachello:1991re,Bijker:1994yr,Giannini:2001kb,Glozman-Riska,Loring:2001kx,Ferretti:2011,Santopinto:2006my,Galata:2012xt,chin,hQM-strange} can reproduce the behavior of observables such as the spectrum and the magnetic moments, but it neglects pair-creation effects, which are manifest as the coupling to meson-baryon (meson-meson) channels.
Above threshold, this coupling leads to strong decays; below threshold, it leads to virtual $qqq - q \bar q$ ($q \bar q - q \bar q$) components in the hadron wave function and shifts of the physical mass with respect to the bare mass \cite{Tornqvist,Horacsek:1986fz,Blask:1987yv,Brack:1987dg,SilvestreBrac:1991pw,Fujiwara:1992yv,Morel:2002vk,Ono:1983rd,Barnes:2007xu,Burns:2012pc}. Some interesting examples of the importance of continuum (or sea) components in the calculation of baryon and meson observables include the self-energy corrections to the $X(3872)$ \cite{Pennington:2007xr,Danilkin:2010cc,charmonium} and $D^*_{s0}(2317)$ \cite{Hwang:2004cd,Rupp:2006sb} bare meson masses, which reconcile QM predictions with the experimental data, and the unquenched quark model calculation of the flavor asymmetry of the nucleon sea \cite{Santopinto:2010zza}.

Nevertheless, the inclusion of continuum effects in the QM in a systematic way is  not an easy task to carry out. In particular, in the baryon sector it is more difficult to do than in the meson one.
Attempts to include continuum effects in the calculation of baryon mass splittings started in the 80's with the work by T\"ornqvist and collaborators \cite{Tornqvist,Ono:1983rd}. Isgur and collaborators showed that the QM emerges as the adiabatic limit of the flux-tube model \cite{Geiger:1996re}; they also proved that the effects of sea pairs on the linear confining potential between a quark and an antiquark is just a renormalization of the string tension \cite{Geiger-Isgur}. 

The self-energy corrections were studied by Horacsek $et$ $al.$ \cite{Horacsek:1986fz}. They calculated the baryon self-energies within a non relativistic Quark Model, considering pion-quark coupling only. 
Specifically, they examined how the self-energies vary among the ground states of the baryon octet and decuplet, and also from ground to excited states.
Brack and Bhaduri \cite{Brack:1987dg} computed the pionic self-energy contributions to the $N$ and $\Delta$ masses perturbatively within the non relativistic QM. 
However,  we shall  show that a complete Pseudo-Nambu-Goldstone boson set should be taken into account. 

Silvestre-Brac and Gignoux \cite{SilvestreBrac:1991pw} studied unitary effects in spin-orbit splittings of $P$-wave baryons. The authors observed that threshold effects can play an important role for the spin-orbit interaction in baryons and that, if these effects are treated correctly, they are capable of explaining the order and importance of spin-orbit splittings in $L=1$ baryons and also partial and total widths.
T\"ornqvist and Zenczykowski \cite{Tornqvist} studied the hadronic mass shifts of the lightest baryons generated by coupled channel effects. Assuming the different contributions from different thresholds to be related by SU(6) symmetry, they derived mass formulas for the relative splittings between $\Delta-N$, $\Sigma^*-\Sigma-\Lambda$ and $\Xi^*-\Xi$ baryons.

A similar study was carried out by Capstick and Morel \cite{Morel:2002vk}, using a pair-creation model for the vertices and Capstick-Isgur model for the bare masses \cite{Capstick:1986bm}. Specifically, they studied the  $\Delta-N$ mass splittings and those of  nonstrange $P$-wave baryons. 

The aim of this article is to present a calculation of the self-energies of ground state octet and decuplet baryons within the unquenched quark model (UQM)  \cite{charmonium,Santopinto:2010zza,bottomonium,bottomonium2,Bijker:2009up,Bijker:2012zza}. 
The UQM is systematic way to include continuum (or loop) effects in the QM. 
These loop corrections are computed considering all accessible ground-state ($1S$) octet and decuplet baryons and a complete set of Pseudo-Nambu-Goldstone bosons.

This is the first step towards a systematic unquenched quark model calculation of the strange and nonstrange baryon spectra with self-energy corrections \cite{BFGS-prep}. After some preliminary studies \cite{Tornqvist,Horacsek:1986fz,Blask:1987yv,Brack:1987dg,SilvestreBrac:1991pw,Fujiwara:1992yv,Morel:2002vk}, several authors have recognized the complexity and importance of this task, which has not been fulfilled yet.

\section{Formalism}
\label{Formalism}
\subsection{Self-energies and continuum components}
We consider  the Hamiltonian
\begin{eqnarray}
H=H_0+V \mbox{ },
\end{eqnarray}
where $H_0$ is the "unperturbed" part, acting only in the bare baryon space, while the second part, $V$ , can couple a baryon state to a continuum made up of baryon-meson intermediate states. We consider $H_0$ free from unitary effects, which  are entirely due to $V$.

In potential model calculations, one considers a certain Hamiltonian $H_0$ and, by fitting the model parameters to the experimental data, one is able to predict the "bare" spectrum of the model.
The bare spectrum is that calculated without continuum effects, namely considering interactions between constituent (valence) quarks only \cite{Eichten:1974af,Isgur:1979be,Godfrey:1985xj,Capstick:1986bm,Iachello:1991re,Bijker:1994yr,Giannini:2001kb,Glozman-Riska,Loring:2001kx,Ferretti:2011,Santopinto:2006my,Galata:2012xt}. 
As widely discussed in the literature, continuum effects may be extremely important, especially in the description of states close to open-flavor (and, sometimes, also hidden-flavor) decay thresholds.

Given this, the physical mass of a baryon, $M_a$, can be written as
\label{eqn:self-system}
\begin{equation}
	\label{eqn:phys-Ma}
	M_a = E_a + \Sigma(E_a)  \mbox{ },
\end{equation}
where $E_a$ is the bare mass and
\begin{equation}
	\label{eqn:self-a}
	\Sigma(E_a) = \sum_{BC} \int_0^{\infty} q^2 dq \mbox{ } \frac{\left| V_{a,bc}(q) \right|^2}{E_a - E_{bc}}  \mbox{ }
\end{equation}
the self-energy correction.
In Eq. (\ref{eqn:self-a}), one has to sum over a complete set of   baryon-meson  intermediate states, $\left| BC \right\rangle$. These channels, with relative momentum $q$ between $B$ and $C$, have quantum numbers $J_{bc}$ and $\ell$ coupled to the total angular momentum of the initial state $\left| A \right\rangle$. $V_{a,bc}$ stands for the coupling between the intermediate state $\left| BC \right\rangle$ and the unperturbed wave function of baryon $A$. $E_{bc} = E_b + E_c$ is the total energy of the channel $BC$, calculated in the rest frame.

This is the formalism used in the present study. Several choices for $V$ are possible. Ours is that of the unquenched quark model of Refs. \cite{charmonium,bottomonium,bottomonium2}. Here, the potential $V$, responsible for the creation of $q \bar q$ pairs, is the $^{3}P_0$ model pair-creation operator, $T^\dag$.

\subsection{An unquenched quark model for baryons}
In the unquenched quark model \cite{charmonium,Santopinto:2010zza,bottomonium,bottomonium2,Bijker:2009up,Bijker:2012zza}, the effects of quark-antiquark pairs are introduced explicitly into the quark model through a QCD-inspired $^{3}P_0$ pair-creation mechanism \cite{3P0}. 
This approach, which is a generalization of the unitarized quark model by T\"ornqvist and Zenczykowski \cite{Tornqvist} (see also Ref. \cite{Geiger:1996re}) is based on a QM, to which $q \bar q$ pairs with vacuum quantum numbers are added as a perturbation and where the pair-creation mechanism is inserted at the quark level. 

Under these assumptions, the baryon wave function is made up of a zeroth order $\left| qqq \right\rangle$ configuration plus a sum over the possible higher Fock components, due to the creation of $^{3}P_0$ $q \bar q$ pairs. To leading order in pair-creation, one has  
\begin{eqnarray} 
	\label{eqn:Psi-A}
	\mid \psi_A \rangle &=& {\cal N} \left[ \mid A \rangle 
	+ \sum_{BC \ell J} \int d \vec{q} \, \mid BC \vec{q} \, \ell J \rangle \right.
	\nonumber\\
	&& \hspace{2cm} \left.  \frac{ \langle BC \vec{q} \, \ell J \mid T^{\dagger} \mid A \rangle } 
	{M_a - E_b - E_c} \right] ~, 
\end{eqnarray}
where 
$A$ is the baryon, $B$ and $C$ represent the intermediate state baryon and meson, $M_a$, $E_b = \sqrt{M_b^2 + q^2}$ and $E_c = \sqrt{M_c^2 + q^2}$ are the corresponding energies, 
$\vec{q}$ and $\ell$ the relative radial momentum and orbital angular momentum between $B$ and $C$, and $\vec{J} = \vec{J}_b + \vec{J}_c + \vec{\ell}$ the total angular momentum.  
The baryon and meson wave functions depend on a single oscillator parameter which, following \cite{Geiger:1996re}, is taken to be $\hbar \omega_{\rm baryon}=0.32$ GeV for the baryons and $\hbar \omega_{\rm meson}=0.40$ GeV for the mesons. 

The $^{3}P_0$ quark-antiquark pair-creation operator, $T^{\dagger}$, is given by  \cite{charmonium,Santopinto:2010zza,bottomonium,bottomonium2,Bijker:2009up,Bijker:2012zza}
\begin{eqnarray}
\label{eqn:Tdag}
T^{\dagger} &=& -3 \, \gamma_0^{\rm eff} \int d \vec{p}_4 \, d \vec{p}_5\ d \vec{P}_\omega \, 
\delta(\vec{p}_4 + \vec{p}_5) \, C_{45}  
{e}^{-\alpha_{d}^2 (\vec{p}_4 - \vec{p}_5)^2/6 }\, 
\nonumber\\
&&F_{45} \,  \Gamma(\vec{P}_\omega)
\left[ \chi_{45} \, \times \, {\cal Y}_{1}(\vec{p}_4 - \vec{p}_5) \right]^{(0)}_0 \, 
b_4^{\dagger}(\vec{p}_4) \, d_5^{\dagger}(\vec{p}_5) ~,   
\label{3p0}
\end{eqnarray}
where $b_4^{\dagger}(\vec{p}_4)$ and $d_5^{\dagger}(\vec{p}_5)$ are the creation operators for a quark and an antiquark with momenta $\vec{p}_4$ and $\vec{p}_5$, respectively.  The function $\Gamma(\vec{P}_\omega)$ represent the flux tube overlap factor, see App. \ref{fluxtube}, that  was introduce by Kokoski  and Isgur \cite{kokoski}. 
The $q \bar q$ pair is characterized by a color singlet wave function, $C_{45}$, a flavor singlet wave function, $F_{45}$, a spin triplet wave function with spin $S=1$, $\chi_{45}$, and a solid spherical harmonic, ${\cal Y}_{1}(\vec{p}_4 - \vec{p}_5)$, which indicates that the quark and antiquark are in a relative $P$-wave. 
Since the operator $T^{\dagger}$ creates a pair of constituent quarks with an effective size, the pair-creation point has to be smeared out by a Gaussian factor, $ \alpha_{\rm d}$. 
$\gamma_0^{\rm eff}$ is an effective pair-creation strength \cite{charmonium,bottomonium,bottomonium2,Kalashnikova:2005ui}, defined as
\begin{equation}
	\label{eqn:gamma0-eff}
	\gamma_0^{\rm eff} = \frac{m_n}{m_i} \mbox{ } \gamma_0  \mbox{ },
\end{equation}
with $i$ = $n$ (i.e. $u$ or $d$) or $s$ (see Table \ref{tab:parameters}). In a recent study \cite{Strong2015}, we discussed the correct treatment of $\gamma_0^{\rm eff}$ in the open-flavor strong decays of baryons. Specifically, we showed that $\gamma_0^{\rm eff}$ can be absorbed in the flavor couplings
\begin{equation}
	\begin{array}{rcl}
	\gamma_0^{\rm eff} \phi_0 & = & \gamma_0^{\rm eff} \frac{1}{\sqrt 3} \left[ |u\bar{u}\rangle + |d\bar{d}\rangle+|s\bar{s}\rangle \right] \\
	& \rightarrow & 	\gamma_0 \phi^{\rm eff}_0 
	= \gamma_0 \frac{|u\bar{u}\rangle +|d\bar{d}\rangle+\frac{m_n}{m_s}|s\bar{s}\rangle}{\sqrt{2+\left(\frac{m_n}{m_s}\right)^2}} 
	\mbox{ }.
	\end{array}
	\label{new3p0}
\end{equation}
$\gamma_0^{\rm eff}$ has then to be fitted to the experimental data, so that a value of $\gamma_0$ can be extracted. The values of the $^3P_0$ pair-creation model parameters are reported in  Table \ref{tab:parameters}. They were  taken from the literature \cite{Santopinto:2010zza,Bijker:2012zza,Geiger:1989yc}, except for the value of the pair-creation strength $\gamma_0^{\rm eff}$, that was fitted to {\color{red}the} strong decay width $\Gamma_{\Delta^{++} \rightarrow p \pi^+} $. See App. \ref{Model parameters} for more details.   

In this paper, we use the operator of Eq. (\ref{eqn:Tdag}) to compute the $^3P_0$ vertices $\left\langle BC \vec q  \, \ell J \right| T^\dag \left| A \right\rangle$, used in the calculation of baryon strong decays and self-energy corrections. Thus, $V_{a,bc}(q) = \sum_{\ell J} \left\langle BC \vec q  \, \ell J \right| T^\dag \left| A \right\rangle$.
The matrix elements of the pair-creation operator $T^{\dagger}$ are derived in explicit form in the harmonic oscillator basis as in Ref. \cite{Strong2015}, using standard Jacobi coordinates. 


\begin{table}[htbp]  
\begin{center}
\begin{tabular}{cc} 
\hline 
\hline \\
Parameter  &  Value    \\ \\
\hline \\
$\gamma_0$ & 17.3     \\  
$\hbar \omega_{\rm baryon}$ & 0.32 GeV \\ 
$\hbar \omega_{\rm meson}$ & 0.40 GeV  \\  
$\alpha_d$      & 0.35 fm  \\
$b$ & 0.18 GeV$^2$  \\
$m_n$      & 0.330 GeV \\
$m_s$      & 0.550 GeV \\ \\
\hline 
\hline
\end{tabular}
\end{center}
\caption{ The values of the $^3P_0$ pair-creation model parameters are taken from the literature \cite{Bijker:2009up,Santopinto:2010zza,Bijker:2012zza,Geiger:1989yc}, except for the value of the pair-creation strength $\gamma_0^{\rm eff}$, that was fitted to strong decay width $\Gamma_{\Delta^{++} \rightarrow p \pi^+} $. }
\label{tab:parameters}  
\end{table}
\section{Self-energy calculation}

\subsection{Mass shifts in the UQM}
\begin{table}[htbp] 
\centering 
\begin{tabular}{cccccc}
\hline
\hline \\
State  &  $J^P$  & $\Sigma(E_a)$ & $E_a$ & $M_a$  \\ \\
$N$     & $\frac{1}{2}^+$            &-0.368  & 1.307&0.939    \\   
$\Lambda$  & $\frac{1}{2}^+$    &-0.465 
&1.581&1.116    \\
$\Sigma$       & $\frac{1}{2}^+$  &-0.303 &1.498
&1.195    \\  
$\Xi$   & $\frac{1}{2}^+$             &-0.474 &1.792 
&  1.318    \\  
$\Delta$  & $\frac{3}{2}^+$         &-0.314 &1.546 
& 1.232     \\  
$\Sigma^*$ & $\frac{3}{2}^+$     &-0.319 &1.702 
&1.383   \\
$\Xi^*$    & $\frac{3}{2}^+$         &-0.345 &1.876
&1.532    \\  
$\Omega$  & $\frac{3}{2}^+$     &-0.352&2.024  
&1.672   \\   \\
\hline
\hline \\
\end{tabular}
\caption{The self energies corrections $\Sigma(E_a)$ (in GeV), calculated with the parameters in Table \ref{tab:parameters}.  The bare energies $E_a$  were obtained in an iterative procedure, to solve the integral equations Eq. (\ref{eqn:phys-Ma}) and (\ref{eqn:self-a}) for each  resonance.  }
\label{resultsold}
\end{table}
Following the formalism of Sec. \ref{Formalism}, we calculate the mass shifts due to continuum effects in the UQM for baryons of Refs. \cite{Santopinto:2010zza,Bijker:2009up,Bijker:2012zza}. 
The procedure is the same as in Ref. \cite{bottomonium}. 

The mass shifts are calculated via Eqs. (\ref{eqn:self-system}), taking the bare energies as free parameters fitted to the reproduction of baryon physical masses. 
If the physical mass of the initial baryon $A$ is above the threshold $BC$, i.e. $M_a > E_b + M_c$, the self energy contribution due to the baryon-meson channel $BC$ is computed as
\begin{equation}
	\label{eqn:SigmaE.real.loops}
	\begin{array}{l}
	\Sigma(E_a)(BC) = \\
	\hspace{0.5cm} \mathcal{P} \int_{M_b+M_c}^{\infty} \frac{dE_{bc}}{E_a - E_{bc}} \mbox{ } 
	\frac{q E_b E_c}{E_{bc}} \left| \left\langle BC \vec q \, \ell J \right| T^\dag \left| A \right\rangle \right|^2 \\
	\hspace{0.5cm} + \mbox{ } 2 \pi i \left\{ \frac{q E_b E_c}{E_a} 
	\left| \left\langle BC \vec q \, \ell J \right| T^\dag \left| A \right\rangle \right|^2 \right\}_{E_{bc} = E_a} 
	\mbox{ },
	\end{array}
\end{equation}
where the symbol $\mathcal{P}$ indicates the Cauchy  principal value of the integral, that is calculated numerically, and the second term,  $2 \pi i \left\{ \frac{q E_b E_c}{E_a} \left| \left\langle BC \vec q \, \ell J \right| T^\dag \left| A \right\rangle \right|^2 \right\}_{E_{bc} = E_a}$ is the imaginary part of the self energy. 

The self-energies $\Sigma(E_a)$, calculated with the $^3P_0$ model parameters of Table \ref{tab:parameters}, are reported in Table \ref{resultsold}. 


\begin{table*}[htbp] 
\centering
\begin{tabular}{ccccccccccc}
\hline
\hline \\
State    & $N \pi$ & $\Sigma K$ & $\Delta \pi$  & $N \eta$ & $N\eta  '$ & $\Delta \eta$ &$\Delta \eta'$& $\Sigma^* K$ &$\Lambda K$ & $\Sigma(E_a)$   \\
\hline \\							
$\Delta$ & -27       &   -14            &     -207 & 0 & 0   &    -36              & -17         &       -13        & 0   & -314    \\ \\   \hline \\							
$N$ & -198      &   -1      &     -138    & -8 & -3   &    0               & 0         &       -7         & -12   & -368  \\ \\
\hline
\hline
\end{tabular}
\caption{Mass shifts (in MeV) of $\Delta\left(\frac{3}{2}^+\right)$ and $N\left(\frac{1}{2}^+\right)$ baryons. The values of the model parameters are given in Table \ref{tab:parameters}.}
\label{tab:Self-delta-01} 
\end{table*}
\begin{table*}[htbp] 
\centering
\begin{tabular}{cccccccccccccc}
\hline
\hline \\
State       & $N \bar K$ & $\Sigma \pi$ & $\Lambda \pi$ & $\Sigma \eta$ &$\Sigma \eta'$                                 &     $\Xi K$ & $\Delta \bar K$ & $\Sigma^* \pi$ &   $\Sigma^* \eta$ &$\Sigma^* \eta'$ & $\Xi^* K$ & $\Sigma(E_a)$ \\
\hline \\							
$\Sigma^*$   & -14       &     -16        &     -15         &            -16  &0            &       -5             &    -99     &     -114 & -7              &-17   &  -17    &  -319   \\    \\
\hline \\							
$\Sigma$   & -7      &     -87       &     -33         &           -18    &-3            &          -22       &   -89       &     -26 &  -14          &0        &  -5      &  -303     \\  
\hline
\hline
\end{tabular}
\caption{As Table \ref{tab:Self-delta-01}, but for $\Sigma^*\left(\frac{3}{2}^+\right)$ and $\Sigma\left(\frac{1}{2}^+\right)$ states.}
\label{tab:Self-Sigma.star} 
\end{table*}

\begin{table*}[htbp] 
\centering
\begin{tabular}{ccccccccccccc}
\hline
\hline \\
State       & $\Sigma \bar K$ & $\Lambda \bar K$ & $\Xi \pi$ &$\Xi \eta'$& $\Xi \eta$ & $\Sigma^* \bar K$ & $\Xi^* \pi$ & $\Xi^* \eta$&$\Xi^* \eta'$  & $\Omega K$ & $\Sigma(E_a)$ \\
\hline \\	
$\Xi^*$ & -35     &     -20        &     -18        &      -23             &0    &       -173          &    -44      &    - 1     &-17       &  -14      & -345    \\    \\
					
\hline \\							
$\Xi$   & -286      &     -9       &     -7        &            -4      &-19      &   -48       &    -64 & -18    &0               &  -20    &  - 474 \\  \\	
\hline
\hline
\end{tabular}
\caption{As Table \ref{tab:Self-delta-01}, but for $\Xi^*\left(\frac{3}{2}^+\right)$ and $\Xi \left(\frac{1}{2}^+\right)$ states.}
\label{tab:Self-Xi.star} 
\end{table*}
The values of the intermediate state baryon and meson masses are taken from the PDG \cite{Nakamura:2010zzi}. 
In Tables \ref{tab:Self-delta-01}--\ref{tab:Self-Lambda}, we show the contributions to the self-energies of ground-state octet and decuplet baryons from the baryon-meson channels we considered in our calculations.

\begin{table}[htbp] 
\centering
\begin{tabular}{ccccccc}
\hline
\hline \\
State       & $\Xi \bar K$ & $\Xi^* \bar K$ & $\Omega \eta$& $\Omega \eta'$ & $\Sigma(E_a)$ \\
\hline \\
$\Omega$ & -120      &     -177        &    -16         &-38        & -352    \\    \\
\hline
\hline
\end{tabular}
\caption{As Table \ref{tab:Self-delta-01}, but for $\Omega\left(\frac{3}{2}^+\right)$ state.}
\label{tab:Self-omega} 
\end{table}

\begin{table*}[htbp] 
\centering
\begin{tabular}{cccccccccccc}
\hline
\hline \\
State       & $N \bar K$ & $\Sigma \pi$ & $\Xi  K$ & $\Lambda \eta$ & $\Lambda \eta'$   &$\Sigma^* \pi$ & $\Xi^* K$ &    $\Sigma(E_a)$ \\
\hline \\							
$\Lambda$   & -180      &     -100        &     -3         &           -6      &-4           &   -156       &     -17    &  -465    \\    \\
\hline
\hline
\end{tabular}
\caption{As Table \ref{tab:Self-delta-01}, but for $\Lambda \left(\frac{1}{2}^+\right)$ state.}
\label{tab:Self-Lambda} 
\end{table*}

\section{Discussion of the results}
\label{Discussion of the results}
We provided the results of a self-consistent calculation of the self-energies of octet and decuplet baryons within the unquenched quark model. 
In the unquenched quark model formalism of Refs. \cite{charmonium,Santopinto:2010zza,bottomonium,bottomonium2,Bijker:2009up,Bijker:2012zza}, the effects of $q \bar q$ sea pairs are introduced explicitly into the QM through a QCD-inspired $^3P_0$ pair-creation mechanism with some modifications.
They include: the use of a quark form factor, as already done by many authors, like T\"ornqvist and Zenczykowski \cite{Tornqvist}, Silvestre-Brac and Gignoux \cite{SilvestreBrac:1991pw} and Geiger and Isgur \cite{Geiger:1996re,Geiger-Isgur}, the introduction of an effective pair-creation strength $\gamma_0^{\rm{eff}}$ \cite{charmonium,bottomonium,bottomonium2,Kalashnikova:2005ui,Strong2015}, and also flux-tube overlap effects \cite{Geiger:1989yc}. 

It is interesting to compare our results to those of previous calculations. In particular, in Table \ref{Comparison} we compare our results with pion and Nambu-Golstone bosons loops with a those  of Ref. \cite{Horacsek:1986fz}. There, the authors only considered pion-quark interaction with a Gaussian form factor with standard deviation  $\Lambda_\pi=516\ $ fm, which is equivalent to $\alpha_{\rm d}=0.9$ fm; as a consequence, they got smaller results for the self-energies, but also a smaller $\Delta-N$ mass splitting.
To complete the discussion, in Table \ref{splitting} we provide a comparison between our $\Delta-N$ mass splitting result with those from Refs. \cite{Horacsek:1986fz,Brack:1987dg}.  In our case, the $\Delta-N$ mass splitting goes from $301$ MeV, when only pion loops are included, to $165$ MeV, when all the Pseudo-Nambu-Goldstone bosons are included. 

\begin{table}
\begin{tabular}{cccc}
\hline
\hline \\
Baryon & UQM & UQM ($\pi$ loops) & HIN \\\\
\hline\\
$\Delta$& -314 &-234&-216 \\\\
N&-368&-336&-198\\\\
$\Sigma^*$&-319&-145&-139\\\\
$\Sigma$&-303&-146&-106\\\\
 $\Xi^*$& -345&-62&-66\\\\
 $\Xi$&-474&-71&-41\\\\
  $\Omega$& -352 &0&--\\\\
 $\Lambda$&-465&-256&-117\\\\
\hline
\hline \\
\end{tabular}
\caption{Comparison between our self-energy calculation results for ground-state octet and decuplet baryons with those from Ref. \cite{Horacsek:1986fz} (HIN). In the second column, we present our results when all Pseudo-Nambu-Goldstone bosons are included. In the third column we present our results when only pion loops are included. Horacsek {\it et al.} (HIN) computed the self-energies only due to pion-quark interaction.} 
\label{Comparison} 
\end{table}

\begin{table}
\begin{tabular}{cccc}
\hline
\hline \\
Model&$\pi$&$\pi K \eta \eta'$&Exp.\cite{Nakamura:2010zzi}\\
\hline \\
UQM&100&54&293\\ \\
HIN&18&-&293\\\\ 
BB&445&-&293\\\\
\hline
\hline \\
\end{tabular}
\caption{$\Delta-N$ mass splitting. In the second column, only pion loops are included. In the third column, all the loops due to Pseudo-Nambu-Goldstone bosons are included. Our results are compared to those by Horacsek {\it et al.} (HIN) \cite{Horacsek:1986fz} and Brack and Bhaduri (BB) \cite{Brack:1987dg}.}
\label{splitting}
\end{table}

The self-energies, we studied in this paper, arise from the coupling to the baryon-meson continuum in the UQM. 
Neglected in naive QM's, these loop effects provide an indication of the quality of the quenched approximation used in QM's calculations, in which only valence quarks are taken into account. 
It is thus worthwhile to see what happens when these pair-creation effects are introduced into the quark model. 
Therefore, we could say that these kind of studies can also be seen as inspections of the QM, of its power in predicting the properties of hadrons and of its range of applicability: if the departure from QM's results is important, one can see new physics emerging or better extra degrees of freedom.  

It has been shown in the meson sector that continuum coupling effects may also be particularly important in the case of suspected non $q \bar q$ states, such as the $X(3872)$ \cite{Choi:2003ue}. 
The uncommon properties of this resonance are due to its proximity to the $D \bar D^*$ decay threshold and can be explained neither within a standard quark-antiquark picture for mesons nor in a simple molecular model. 
In Ref. \cite{charmonium}, it is shown that the continuum coupling effects of the $X(3872)$ can give rise to $D\bar D^*$ and $D^*\bar D^*$ components in addition to the $c \bar c$ core and determine a downward energy shift, which is necessary to obtain a better reproduction of the experimental data.
In other words, while far from thresholds loop effects can be partially re-absorbed into a new set of renormalized parameters for the potential \cite{charmonium,Geiger:1989yc}, this is not possible for those states that lie close to a meson-meson decay threshold. There are also suspected non $qqq$ baryons, like the $\Lambda^*(1405)$, which may deserve an inspection within the UQM formalism.


\begin{appendix}

\section{Parameters of the $^3P_0$ pair-creation model}
\label{Model parameters}
The values of the $^3P_0$ pair-creation model parameters are taken from the literature \cite{Santopinto:2010zza,Bijker:2009up,Bijker:2012zza,Geiger:1989yc}, except for the value of the pair-creation strength $\gamma_0^{\rm eff}$ (see Table \ref{tab:parameters}), that has to be fitted to the reproduction of experimental strong decay widths. 
We have chosen to fit $\gamma_0^{\rm eff}$ to the experimental strong decay width $\Delta^{++} \rightarrow p \pi^+$ \cite{Nakamura:2010zzi}.
In this case, since the created pair $q \bar q$ is $d \bar d$, the effective pair creation strength $\gamma_0^{\rm eff}$ coincides with $\gamma_0$ [see Eq. (\ref{eqn:gamma0-eff})].
 
The decay width is calculated within the $^3P_0$ model \cite{SilvestreBrac:1991pw,Strong2015,Capstick:1992th} as 
\begin{equation}
	\begin{array}{rcl}
	\Gamma_{\Delta^{++} \rightarrow p \pi^+} & = & 2 \Phi_{A \rightarrow BC} \left| \left\langle BC 
	\vec q_0  \, \ell J \right| T^\dag \left| A \right\rangle \right|^2  \\ 
	& = & \Phi_{\Delta^{++} \rightarrow p \pi^+} \\ 
	& & \left| \left\langle p \pi^+
	\vec q_0  \, 1 \frac{1}{2} \right| T^\dag \left| \Delta^{++} \right\rangle \right|^2 \\ 
	\end{array}
\end{equation}
where $\left\langle BC \vec q_0  \, \ell J \right| T^\dag \left| A \right\rangle$ is a $^3P_0$ amplitude, describing the coupling between the baryon $\left| A \right\rangle = \left| \Delta^{++} \right\rangle$ and the final state $\left| BC \right\rangle = \left| p \pi^+ \right\rangle$, and 
\begin{equation}
	\Phi_{A \rightarrow BC} = 2 \pi q_0 \frac{\tilde M_b \tilde M_c}{M_a}  
\end{equation}
is Capstick and Roberts' effective  phase space factor \cite{Capstick:1992th}, where $\tilde M_b = \tilde M_p = 1.1$ GeV and $\tilde M_c = \tilde M_\pi = 0.72$ GeV are the masses of the hadrons $B$ and $C$ in the weak-binding limit. 

\section{Flux-tube breaking model}
\label{fluxtube}
\begin{figure}[htb]
\centering
   \includegraphics[scale=0.5]{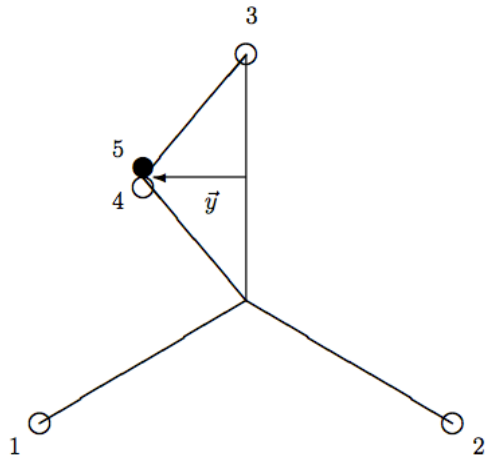}\caption[]{\small $A \rightarrow BC$ baryon decays by flux-tube breaking. 
Open circles denote quarks and closed circles antiquarks.} 
\label{baryon}
\end{figure}

Finally, the function $\Gamma(\vec P_w)$ indicates the flux-tube overlap function in the momentum space, that describes the overlap of the flux-tube of the initial particle with those of the final ones. This mechanism was originally developed by Kokoski and Isgur for string-like meson decays in Ref. \cite{kokoski} as an extension of the $^3P_0$ model of hadron decays \cite{3P0} (see also Refs. \cite{Geiger:1989yc,Blundell:1995ev,Page:1995rh}).
In the flux-tube breaking model, it is assumed that the flux-tube breaking takes place at a distance $\vec{y}=\vec{r}_4- (\vec{r}_1+\vec{r}_2+\vec{r}_3)/3= -\vec{x}-\sqrt{\frac{3}{2}\vec{\lambda}}$ from the center of mass of the initial baryon $A$ (see figure \ref{baryon}). 
The modification of the transition amplitude consists in the addition of a flux-tube overlap function $\gamma(\vec{y}) $ which, just as for the mesons, is taken to be a Gaussian
\ba
\gamma(\vec{y}) = \mbox{e}^{-b \vec{y}^{\, 2}/2} ,
\ea
in coordinate space, and its  Fourier transform $\Gamma(\vec{P}_\omega)$ is the flux-tube overlap in momentum space
\ba
\Gamma(\vec{P}_\omega) = \frac{1}{(2\pi)^{3/2}} \, \int d \vec{y} \, 
\mbox{e}^{-i \vec{y} \cdot \vec{P}_\omega} \, \gamma(\vec{y}) 
= \frac{\mbox{e}^{-\vec{P}_\omega^{\, 2}/2b} }{b^{3/2}} , 
\ea 
with $\vec{P}_\omega$  the canonic conjugate coordinate to the vector $\vec{y}$.

Note, that the Gaussian form is similar, but the coordinates  are different  to those used by Stancu-Stassart \cite{stancu} and Geiger-Isgur \cite{Geiger:1996re,Geiger-Isgur}.
 
 The UQM contains the $^3P_0$ transition  amplitude, it   is written in terms of a spherical basis, but it can be easily expressed in a plane-wave basis  by the change of the variables
\begin{eqnarray}
\begin{array}{c}
  \vec{K}=\vec{p}_{_{B}}+\vec{p}_{_{C}}\\
 \vec{q}=\frac{2}{5}\vec{p}_{_{B}}-\frac{3}{5}\vec{p}_{_{C}}
\end{array} \Longleftrightarrow 
\begin{array}{c}
 \vec{p}_{_{B}}=\frac{3}{5}\vec{K}+ \vec{q}\\
\vec{p}_{_{C}}=\frac{2}{5}\vec{K}- \vec{q}
\end{array} \label{changecoordinate}
\end{eqnarray}
Thus transition amplitude can be  written as 
\begin{eqnarray}
 \langle
BC,J_b, m_b,J_c, m_c,\vec{p}_B,\vec{p}_C|T^{\dagger}|A,\vec{P}_A,J_a,m_a
\rangle \nonumber  \\= 3\gamma\sum_{m}\langle 1,1;m,-m|0,0\rangle
\nonumber\langle\Phi^{m_b}_B\Phi^{m_c}_C|\Phi^{m_a}_A\Phi^{-m}_
{vac}\rangle  \\ \times I_m(A\rightarrow BC) ,\label{amp3p0s} \ \ \ \ \ \ \ \ \ \ \ \ \ \ \ \
\end{eqnarray}
where 
\begin{eqnarray}
 \Phi^{-m}_
{vac}= \chi^{-m}_1\phi_0.
\end{eqnarray} is the spin-flavor wave function of the pair  created from the  vacuum in a triplet state $S=1$, $\langle\Phi^{m_b}_B\Phi^{m_c}_C|\Phi^{m_a}_A\Phi^{-m}_
{vac}\rangle $ correspond to the flavor-spin overlap, and the spatial overlap is $ I_m(A\rightarrow BC)$.

The  spatial-transition  amplitude in the flux-tube breaking model is given by 
\ba
I_m(A \rightarrow BC) &=& \sqrt{\frac{3}{4\pi}} \, \delta(\vec{p}_B+\vec{p}_C) \, 
\frac{1}{(2\pi)^{3/2}} \, \int d \vec{r} \, d \vec{x} \, \nonumber \\ &\times&\gamma(\frac{\vec{r}}{2}-\vec{x}) \, 
\mbox{e}^{-i \vec{r} \cdot \vec{p}_B/2}  \psi_B^{\ast}(\vec{r}-\vec{x}) 
\nonumber\\
&\times&\, \psi_C^{\ast}(\vec{x}) 
\vec{\epsilon}_m \cdot (-2i\vec{\nabla}_r-2\vec{p}_B) \, \psi_A(\vec{r}) 
\label{mspace2}
\ea
in coordinate space, and by 

\ba
I_m(A \rightarrow BC) &=& \frac{-2}{3} \, \sqrt{\frac{1}{4\pi}} \, 
\, \frac{\delta(\vec{p}_B+\vec{p}_C) }{(2\pi)^{3/2}} \, 
\int d \vec{p}_{\rho} \, d \vec{p}_{\lambda} \, d \vec{P}_\omega \, 
  \nonumber \\   &\times&\Gamma(\vec{P}_\omega) \, \vec{\epsilon}_m \cdot (\vec{p}_C+\sqrt{\frac{2}{3}} \vec{p}_{\lambda}) \,\phi_A(\vec{p}_{\rho},\vec{p}_{\lambda})
\nonumber\\
&\times& \phi_B^{\ast}(\vec{p}_{\rho},
\vec{p}_{\lambda}+\sqrt{\frac{2}{3}} \vec{p}_C - \sqrt{\frac{1}{6}} \vec{P}_\omega) \, 
  \nonumber \\   &\times&
\phi_C^{\ast}(-\frac{\vec{p}_C}{2}-\sqrt{\frac{2}{3}} \vec{p}_{\lambda} - \frac{2 \vec{P}_\omega}{3})  
\label{bmom2}
\ea
in momentum space, with \begin{eqnarray}
\nonumber \vec{p}_{_{\rho}}&=&\frac{1}{\sqrt{2}}\left(\vec{p}_{_{1}}-\vec{p}_{_{2}} \right) 
\label{prho} \\ \nonumber
\vec{p}_{_{\lambda}}&=& \frac{1}{\sqrt{6}}\left(\vec{p}_{_{1}}+\vec{p}_{_{2}}-
2\vec{p}_{_{3}}  \right) \label{plamda}.
\end{eqnarray}

\subsection{Pair creation vertex}

Finally, the smearing of the pair creation vertex can be included in momentum space by adding 
an exponential factor to give 
\ba
I_m(A \rightarrow BC) &=& \frac{-2}{3} \, \sqrt{\frac{1}{4\pi}} \, 
  \frac{\delta(\vec{p}_B+\vec{p}_C)}{(2\pi)^{3/2}} \int d \vec{p}_{\rho} \, d \vec{p}_{\lambda} \, d \vec{P}_\omega\nonumber \\  &\times&
 \, \Gamma(\vec{P}_\omega) \, 
\vec{\epsilon}_m \cdot (\vec{p}_C+\sqrt{\frac{2}{3}} \vec{p}_{\lambda}) \,
\nonumber\\
&\times& \phi_B^{\ast}(\vec{p}_{\rho},
\vec{p}_{\lambda}+\sqrt{\frac{2}{3}} \vec{p}_C - \sqrt{\frac{1}{6}} \vec{P}_\omega ) \, 
 \nonumber \\  &\times&
\phi_C^{\ast}(-\frac{\vec{p}_C}{2}-\sqrt{\frac{2}{3}} \vec{p}_{\lambda} - \frac{2}{3}\vec{P}_\omega) \, 
\nonumber\\
&\times&  \phi_A(\vec{p}_{\rho},\vec{p}_{\lambda}) \,  
\mbox{e}^{-2r_q^2 (\vec{p}_C+\sqrt{\frac{2}{3}} \vec{p}_{\lambda})^2/3} .
\label{bmom3}
\ea

\end{appendix}


\end{document}